\documentclass[traditabstract]{aa} 
\usepackage[pdftex]{graphicx}
\usepackage{txfonts}
\usepackage{natbib}
\usepackage{longtable}

\def\m2s2{\hbox{\,m$^{2}$\,s$^{-2}$}} 
\def\sini{\hbox{sin\,$i$}}      
\def\Msun{\hbox{$M_{\odot}$}}             
\def\Rsun{\hbox{$R_{\odot}$}}

\begin{document}

\title{Detection of Neptune-size planetary candidates with CoRoT data}
\subtitle{Comparison with the planet occurrence rate derived from Kepler}

    \titlerunning{Detection of small-size planetary candidates with CoRoT data}
    \authorrunning{A. S. Bonomo et al. 2012}

\author{A.~S.~Bonomo \inst{1,2} 
\and P.~Y.~Chabaud \inst{1} 
\and M.~Deleuil\inst{1}
\and C.~Moutou\inst{1}
\and F.~Bouchy\inst{3,4}
\and J.~Cabrera\inst{5}
\and A.~F.~Lanza\inst{6}
\and T.~Mazeh\inst{7} 
\and S.~Aigrain\inst{8} 
\and R.~Alonso\inst{9}
\and P.~Guterman\inst{1}
\and A.~Santerne\inst{1, 4}
\and J.~Schneider\inst{10} 
}

\institute{
Laboratoire d'Astrophysique de Marseille, Universit\'e Aix-Marseille \& CNRS, 38 rue Fr\'ed\'eric Joliot-Curie, F-13388 Marseille Cedex 13, France
\and INAF - Osservatorio Astronomico di Torino, via Osservatorio 20, 10025 Pino Torinese, Italy
\and Institut d'Astrophysique de Paris, UMR7095 CNRS, Universit\'e Pierre \& Marie Curie, 98bis boulevard Arago, 75014 Paris, France
\and Observatoire de Haute-Provence, Universit\'e Aix-Marseille \& CNRS, F-04870 St.~Michel l'Observatoire, France 
\and Institute of Planetary Research, German Aerospace Centre, Rutherfordstrasse 2, 12489 Berlin, Germany
\and INAF - Osservatorio Astrofisico di Catania, Via S. Sofia, 78, 95123 Catania, Italy
\and School of Physics and Astronomy, Raymond and Beverly Sackler Faculty of Exact Sciences, Tel Aviv University, Tel Aviv, Israel 
\and Oxford Astrophysics, Denys Wilkinson Building, Keble Road, Oxford OX1 3RH, UK
\and Observatoire de l'Universit\'e de Gen\`eve, 51 chemin des Maillettes, 1290 Sauverny, Switzerland 
\and LUTH, Observatoire de Paris, CNRS, Universit\'e Paris Diderot; 5 place Jules Janssen, 92195 Meudon, France
}


	

\offprints{A.~S.~Bonomo\\
\email{aldo.bonomo@oamp.fr}}

\date{Received June 8, 2012; accepted September 13, 2012}

\abstract{
\emph{Context.} The CoRoT space mission has been searching for transiting planets 
since the end of December 2006.
It has led to the detection of about twenty Jupiter-size planets and three planets with radius
$R_{\rm p}\lesssim5~R_{\oplus}$. The latter are CoRoT-7b, 
the first super-Earth observed in transit, and two validated Neptunes,
CoRoT-24b and c, in a multiple system. \\
\emph{Aims.} We aim to investigate the capability of CoRoT to detect small-size transiting
planets in short-period orbits, and to compare
the number of CoRoT planets with $2.0 \leq R_{\rm p} \leq 4.0~R_{\oplus}$ with the
occurrence rate of small-size planets provided by the distribution of Kepler planetary candidates
(Howard et al. 2012, ApJS, 201, 15). \\
\emph{Methods.} We performed a test that simulates transits of super-Earths and Neptunes in real CoRoT light curves
of six long observational runs and searches for them blindly by using the transit detection pipeline
developed at the Laboratoire d'Astrophysique de Marseille. \\
\emph{Results.} The CoRoT detection rate of planets with radius between 2 and $4~R_{\oplus}$ 
and orbital period $P \leq 20$~days is 59\% (31\%) around stars brighter 
than $r'=14.0$ (15.5). The vast majority of the missed planets went 
undetected because of a low transit signal-to-noise ratio ($S/N$).
However, in some cases, additional instrumental or astrophysical noise may prevent
even transits with relatively high $S/N$, i.e. $S/N \geq 10$, from being revealed.
By properly taking the CoRoT detection rate for Neptune-size planets
($2 \leq R_{\rm p} \leq 4~R_{\oplus}$) and the transit probability into account, 
we found that according to the Kepler planet 
occurrence rate, CoRoT should have 
discovered $12 \pm 2$ Neptunes
orbiting G and K dwarfs with $P \leq 17$~days in six observational runs. 
This estimate must be compared with the validated Neptune CoRoT-24b and 
five CoRoT planetary candidates in the considered range of planetary radii, the nature of which 
is still unsolved. We thus found a disagreement with 
expectations from Kepler at 3~$\sigma$ or 5~$\sigma$, assuming
a blend fraction of 0\% (six Neptunes) and 100\% (one Neptune) for these candidates. \\
\emph{Conclusions.} This underabundance of CoRoT Neptunes with respect to Kepler 
may be due to  
\emph{a)} an underestimate of the uncertainty on the Kepler 
planet occurrence;
\emph{b)} an underestimate 
of the false-positive probability of the Kepler small-size 
planetary candidates;
\emph{c)} an overestimate of our prediction of CoRoT Neptunes related
to the number of G and K dwarfs observed by CoRoT; or
\emph{d)} different stellar populations probed by the two space missions.
Regardless of the origin of the disagreement, which needs to be investigated in more detail,  
the noticeable deficiency of CoRoT Neptunes at short orbital periods seems to indirectly support 
the general trend found in Kepler data,
i.e. that the frequency of small-size planets increases with increasing orbital 
periods and decreasing planet radii.}

	
  
\maketitle
%

\section{Introduction}
More than 750 exoplanets have been found since the discovery of the
extrasolar planet 51~Peg~b around a solar-like
star by \citet{MayorQueloz95}.
Most of them are gaseous giant planets, but much effort is 
currently put into detecting small-size
planets, such as Neptunes, super-Earths, and Earth-sized planets, 
with radial-velocity and space-based transit surveys.
 
About a hundred low-mass planets have been recently 
discovered by the HARPS and HIRES radial-velocity surveys (e.g., \citealt{Mayoretal11}, 
\citealt{Howardetal10}, and references therein).
From these early discoveries, low-mass planets seem to be 
abundant. Focusing on 376 solar-type stars (from late-F to late-K dwarfs) 
observed with the HARPS spectrograph, \citet{Mayoretal11} found 
that about $28 \pm 5\%$ of them host a planet with a 
minimum mass $M \sini=3$~to 30 M$_\oplus$ and 
orbital period $P<50$~days. 
With a more limited sample of G and K dwarfs observed 
with the HIRES spectrograph,
\citet{Howardetal10} reported an occurrence rate of $18 \pm 5$\%,
slightly lower than the rate of \citet{Mayoretal11}, for the same 
mass domain and period range. 

The observed occurrence rate of low-mass planets around dwarf stars,
its dependence on the properties of the host stars (metallicity,
stellar mass, etc.) and on the planetary parameters (mass, semi-major axis, 
and eccentricity) represent a powerful test for models of planet 
formation and migration (e.g., \citealt{Mordasinietal09a}, \citealt{Mordasinietal09b}, 
\citealt{Mordasinietal12} and references therein; \citealt{Mayoretal11}). 

In addition to their contribution to general statistics, transiting low-mass planets  
are extremely interesting to study because one can measure their radius and mass
free of the $\sin{i}$ degeneracy if radial-velocity variations induced by the planet
on its parent star or transit timing variations in a multiple system are observed. 
This enables a detailed characterisation of small-size planets and studies of 
their internal structure.

The transits of small-size planets can be observed from the ground
only if their host stars have sizes considerably smaller than that of the Sun.
This is, for example, the case of the super-Earth or 
mini-Neptune GJ~1214b \citep{Charbonneauetal09}. 
In general, we have to move to space to detect small-size planets around solar-like stars
as the photometric precision from the ground
is insufficient. 

\subsection{The CoRoT space mission.}
The CoRoT space mission is the pioneer for 
the detection of transiting planets 
from space. It is operated by CNES\footnote{French space agency: 
Centre National 
d'\'Etudes Spatiales}, 
with contributions from Austria, Belgium, Brazil, ESA, Germany and Spain
\citep{Baglin03}.
CoRoT was launched at the end of 
December 2006 on a polar orbit at an altitude of $\sim 900$~km. With its 
27~cm telescope, it has been monitoring the optical flux of thousands of stars with
$11<V<17$ for  $\sim150$ days at the longest, in several fields of view 
pointing mainly in two different directions:
one towards the Aquila and the other one towards the Monoceros constellation.
The field of view of the telescope for the exoplanet science 
was $\sim 4~\rm{deg}^{2}$ before
one DPU on board was lost through technical problems on 8~March 
2009. Since then, only one 
CCD dedicated to exoplanet science is currently working, giving a 
halved field of view of $1.38~\rm{deg}^{2}$. We refer the reader 
to \citet{Auvergneetal09} for a detailed description of the instrument.

To date, CoRoT has led to the detection of three small-size
planets with $ R_{\rm{p}}  \lesssim 5~R_{\oplus}$ : CoRoT-7b, the first  
super-Earth observed in transit \citep{Legeretal09}, and 
two transiting Neptunes, CoRoT-24b and CoRoT-24c, in a multiplanet 
system \citep{Alonsoetalsubm}. For the last two planets, radial-velocity 
observations performed with the HARPS spectrograph did 
not have sufficient precision to derive their masses, given that 
their parent star is faint. However, by using a 
procedure similar to that described in 
\citet{Torresetal11}, these candidates 
have been ``validated'' as low-mass 
planets \citep{Alonsoetalsubm}. This means that
the probability that the observed transits are due 
to astrophysical false positives, such as blended eclipsing binaries, 
is very low. In other words, the planet scenario is most likely 
true \citep{Torresetal11}.

\subsection{The Kepler space mission.}
The other space-based mission aimed at detecting 
planetary transits is the NASA Kepler mission, launched in April 2009
in a heliocentric orbit. It performs photometric observations
of $\sim 156,000$ stars with $9<V<16$ in a fixed 
stellar field of $115~\rm{deg}^{2}$ located in the 
Cygnus constellation \citep{Boruckietal06, Kochetal10}. 
With its 95~cm telescope and 42~back-illuminated CCDs, it reaches
a photometric precision of $\sim 50$~ppm on a $m_{\rm V}=11.5$ star 
for a $\sim 30$~min integration time, while the corresponding 
RMS of CoRoT is $\sim 200$~ppm.
Moreover, since Kepler has been observing always the same stellar field,
as time goes by, it has the capability of detecting 
shallower and shallower transits because the  
signal-to-noise ratio ($S/N$) scales as the square root of the number of transits.
Thanks to the unprecedented photometric precision of the instrument,
the Kepler team found 2321 planetary candidates in the first sixteen months 
of data \citep{Batalhaetal12}. The vast majority of them 
have radii below the Neptune radius. 

Up to now, the Kepler team has announced the discovery of  
42 planets with radius smaller than $5~R_{\oplus}$, 
many of which belong to multiple 
systems\footnote{http://kepler.nasa.gov/Mission/discoveries/}. We recall the exceptional 
discoveries of the Kepler-11 and Kepler-20 systems 
with six and five small-size planets, respectively. In many cases, 
for instance Kepler-20, the planets have been only ``validated'' because 
their mass has not been measured 
either through radial-velocity observations or
transit timing variations (e.g., \citealt{Torresetal11, Fressinetal11a}).
Kepler-20e with $R_{\rm p}=0.87~R_{\oplus}$ is the smallest 
planet found by the Kepler team \citep{Fressinetal11b}.

From stellar population synthesis and
Galactic structure models, \citet{MortonJohnson11} claimed that
more than 90\% of Kepler planetary candidates are very likely 
real planets. Based on this fiducial lower limit, 
\citet{Howardetal12} have performed detailed statistical studies on the Kepler 
planetary candidates. After correcting for the alignment 
probability, these authors derived an occurrence rate of 
$13 \pm 1 \%$ for planets with radius $2 \leq R_{\rm p} \leq 4~R_{\oplus}$, 
orbiting G and K dwarfs in $P < 50$~days. 
Moreover, Howard and coworkers found that the planet occurrence increases 
with decreasing planet radius and increasing orbital 
period.

\subsection{Outline and purpose of the present work}
In the present work, we first of all aim to investigate the capability of CoRoT
of detecting small-size planetary transits. To that purpose, we 
simulated planetary transits of super-Earths and Neptunes in real
CoRoT light curves of G and K dwarfs and searched for them blindly, that is, 
without knowing in advance in which light curves 
the transits had been simulated (Sect.~\ref{bt5}). We used the transit detection
pipeline currently operating at the Laboratoire d'Astrophysique de 
Marseille (Sect.~\ref{data_analysis}), which is in charge of 
the CoRoT alarm mode. This allowed us 
to derive the CoRoT detection rate for planets with
different sizes, from 1.3 to $5.0~R_{\oplus}$ (Sect.~\ref{corot_det_rate}), and 
to investigate the obstacles to the detection of small-size planets with 
CoRoT (Sect.~\ref{det_obstacles}). By deriving the CoRoT detection 
rate of planets with radii from 2.0 to $4.0~R_{\oplus}$, 
we compared the number of small-size planets discovered by CoRoT 
with expectations from the Kepler occurrence rate provided by \citet{Howardetal12}, 
after taking the transit probability into account (Sect.~\ref{comp_kepler}).

\section{Simulations}
\label{bt5}
We made use of real light curves (N2 level data) of the
CoRoT long runs lasting more than 110~days, i.e. 
LRa01 \citep{Caroneetal12}, LRa02, LRa03 \citep{Cavarrocetal12}, LRa04, 
LRc01 \citep{Cabreraetal09}, and LRc02 \footnote{data available at
http://idoc-corot.ias.u-psud.fr/}. They refer to six different 
fields of view observed by CoRoT.
For all runs except for LRa03 and LRa04, both of the two CCDs dedicated
to exoplanet science were operating. 
Out of the $\sim 55 \, 000$ stars observed 
in the above-mentioned long runs, almost $14 \, 000$ were classified 
as dwarfs with spectral type G and K,
according to the photometric classification of 
Exodat \citep{Deleuiletal09}. The remaining targets 
were ranked as either evolved or hotter stars.
Among the $\sim 14 \, 000$ G and K dwarfs,
only those brighter than $r'=15.5$ were selected, that is, 9526 stars.
From this sample, light curves of binaries or
planetary candidates found by the CoRoT detection team were excluded.
This yields 9333 light curves for our blind test. 

Here we must point out that from a comparison
between the Exodat photometric classification and a 
massive spectral analysis of more 
than $1\,000$ CoRoT stars with 
\emph{FLAMES/GIRAFFE} multi-fiber observations, 
\citet{Gazzanoetal10} found that the former
tends to systematically underestimate the number of dwarfs, while
there is a fairly good agreement between the spectral types determined
by the two methods.
Figure~8 in \citet{Gazzanoetal10} clearly shows this behaviour for the 
LRa01 and LRc01 stellar fields: the spectral analysis found 
52\% of the dwarfs while only 32\% of the stars were classified as such by
the photometric classification, which is a difference of a factor 1.7.
The number of subgiants identified by both methods
is similar, while the amount of giants was 
clearly overestimated by Exodat.
However, since the classification based on spectroscopy 
is only available for about a thousand targets
in the LRa01 and LRc01, we could only rely on the
Exodat classification for the light curve selection.
Nevertheless, we will consider a correction 
factor to take the underestimate of dwarf stars 
by the Exodat classification
into account in Sect.~\ref{comp_kepler}.

The choice of restricting ourselves to the light curves of G and K dwarfs 
allows us to compare the rate of 
small-size planets detected by CoRoT with the statistics provided by the 
distribution of Kepler planetary candidates.
In addition, G and K main-sequence stars 
with $r' \leq 15.5$ are the most interesting targets
for the detection of small-size planets.
The fainter the stars and/or the bigger their radius, the more difficult 
the discovery of Neptunes and super-Earths. 

Simulated transits of small-size planets, super-Earths and Neptunes, with radius $R_{\rm p}$ 
ranging between 1.3 and 5.0~$R_{\oplus}$ and orbital period $P$ 
from 0.7 up to 20~days were inserted into the selected 9333 CoRoT 
light curves. A flat distribution was adopted both for the planetary radius
and the orbital period. 
To obtain a sufficiently large statistics
for the detection of super-Earths, we repeated our experiment twice. 
First we simulated only transits of
planets with $1.3 <R_{\rm p} \leq 2.5~R_{\oplus}$ in the 9333 light curves.
Then, we inserted transits of bigger planets 
with $2.5 < R_{\rm p} \leq 5.0~R_{\oplus}$ into the same original light curves. 

We used the universal transit modeler \citep{Deeg09} to simulate 
the planetary transits assuming, for simplicity, circular planetary orbits, 
the stellar radius and mass of a G2V star for the G dwarfs, i.e. $R_{\star}=1~\Rsun$ 
and~$M_{\star}=1~\Msun$, and those of a K2V star for the K ones:
$R_{\star}=0.8~\Rsun$ and $M_{\star}=0.7~\Msun$ \citep{Cox00}.
A linear limb-darkening law was adopted with the linear coefficient 
for the CoRoT bandpass $u=0.62$ or 0.69 for the G or K spectral types, 
respectively \citep{Sing10}. The impact parameter was chosen at random
between 0 and 1 to take the distribution of the orbital inclination into account.
The initial epoch of the transits was also chosen randomly.

Our simulations do not take planets around 
one component of a binary or triple physical system into account. 
Such planets would be more difficult to detect because
their transit depth is diluted by the stellar companion(s). However, for the time 
being, it is not clear how the frequency of small-size planets is affected 
by the multiplicity of stellar systems and as long as we lack 
this information, we will consider only single stars. 
Moreover, even the distribution of the planetary radii of Kepler candidates,
which was used by \citet{Howardetal12} to determine the planet occurrence as a 
function of orbital period, implicitly assumes that they orbit
single stars \citep{Boruckietal11, Batalhaetal12}.

Another possible cause of dilution of the transit depth is the contamination
of background stars inside the CoRoT aperture mask of a star \citep{Deegetal09}. 
However, this can be readily neglected for our study,
because this contamination is typically only $\sim 5\%$ of the flux collected by CoRoT
for a given star (e.g., \citealt{Bonomoetal10} ).

\section{Data analysis} 
\label{data_analysis}
\subsection{The LAM transit detection pipeline}
\label{lam_pipeline}
Transits were searched for by means of the transit detection
pipeline developed at the Laboratoire d'Astrophysique de Marseille (LAM). 
The CoRoT detection team uses seven different techniques to 
detrend the light curves and search for transits. Blind tests organized within the
CoRoT consortium to compare their performance on real CoRoT 
light curves have shown that the LAM pipeline 
and two other algorithms have the best performance and
a similar detection rate (P. Barge, priv. comm.).
Our pipeline consists of eight steps:

a) Outliers caused by proton impacts during the 
passage of the satellite across the South Atlantic Anomaly 
\citep{Auvergneetal09} are filtered out with a 5-sigma clipping.

b) Data points oversampled at 32~s are re-binned at 512~s.

c) Low-frequency variations (i.e., stellar variability) 
are removed with a high-pass filter, consisting of a sliding median
filter with a window extension 
of 0.5 days (see Sect.~\ref{window_median_filter} for a 
discussion on the choice of the filtering time scale).

d) High-frequency variations are removed with a low-pass filter, 
consisting of a Savitzky-Golay filter \citep{Pressetal92} 
with a time scale of $\sim1$~hour.
Most of these high-frequency variations are related to the
signal at the satellite orbit of 6184~s (1.72~h).

e) Jumps/discontinuities that are mainly caused by 
hot pixels (\citealt{Srouretal03}, cf. \citealt{Auvergneetal09} for the specific 
case of CoRoT) or pointing displacements are automatically detected.
After that, data points that fall 0.4~days before and after 
the time of a given hot pixel are removed. These discontinuities 
in CoRoT light curves may represent a major concern for detecting  
shallow planetary transits (see below Sect.~\ref{det_obstacles}). Indeed, after the filtering 
process (steps a-b-c-d), they give rise
to dips in the residuals that can be erroneously identified as transits
by the detection algorithms, such as the box fitting least squares 
(BLS, \citealt{Kovacsetal02}). 

\begin{figure}[h!]
\centering
\vspace{-1cm}
\includegraphics[width=10cm, angle=180]{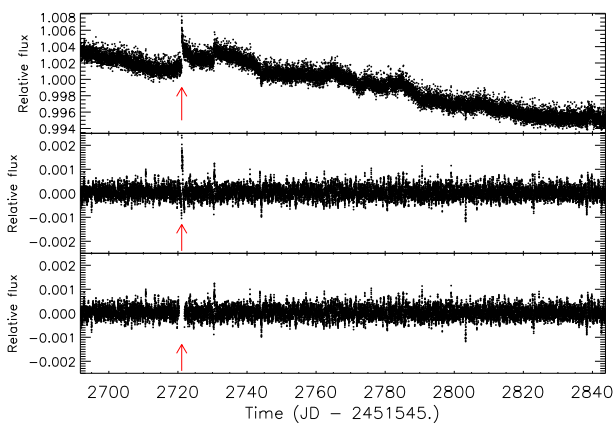}
\vspace{-0.5cm}
\caption{\emph{Top panel}: the CoRoT light curve LRc01\_E1\_0331,
filtered from cosmic rays (Sect.~\ref{lam_pipeline}.a),
containing simulated transits of a super-Earth with an orbital period 
of 9.8~days. The red arrow
indicates the jump produced by a hot pixel and correctly
identified by our pipeline.
\emph{Middle panel}: high- and low-pass filtered light curve
after steps b), c), and d). Planetary transits by the super-Earth 
are not visible to the naked eye. Other artificial dips 
are produced by undetected small discontinuities. 
\emph{Bottom panel}: light curve after removal of data points 
in the vicinity of the identified discontinuity (Sect.~\ref{lam_pipeline}.e).
}
\label{lc_331}
\end{figure}

The detection of hot pixels is carried out
by using a sliding window of eight data points sampled at 512~s 
and computing the standard error of the mean
inside the moving window. 
When a discontinuity
takes place in the light curve, the standard error  
suddenly increases, allowing the hot pixel detection as soon 
as it exceeds a given threshold (see Fig.~\ref{lc_331}). 
The latter was fixed at four standard deviations
of the set of the mean standard errors
obtained for each position of the 
eight-point sliding window. 
This method allows us to detect approximately
60-70\% of the most evident hot pixels. After detecting 
a discontinuity in the light curve,
we prefer to remove data points in its vicinity. Indeed, after a simple adjustment
of the light curve without any clearing of data points
in the vicinity of a hot pixel, high-frequency variations could still remain
and hinder the transit detection.

\begin{figure}[h!]
\centering
\vspace{-1cm}
\includegraphics[width=10cm, angle=180]{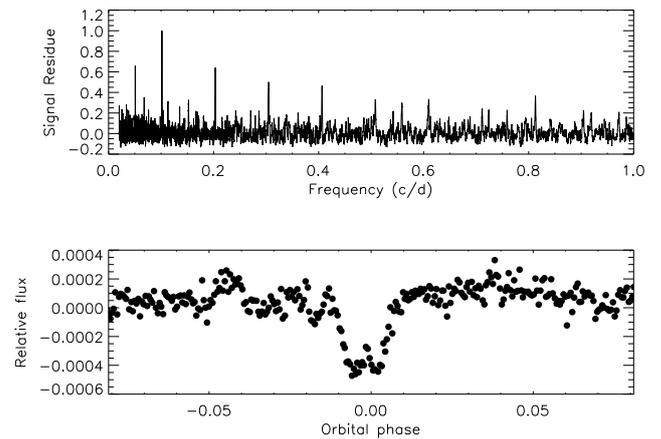}
\vspace{-0.5cm}
\caption{\emph{Top panel}: BLS spectrum of the filtered light curve LRc01\_E1\_0331,
containing simulated transits by a super-Earth with orbital period $P=9.8$~days (see
bottom panel of Fig.~\ref{lc_331}). The highest peak corresponds to the
inserted transit signal, which is thus easily detected. Its harmonics are also clearly
visible. 
\emph{Bottom panel}: phase-folded light curve at the period and epoch found by the
BLS (Sect.~\ref{lam_pipeline}.g).
}
\label{bls_331}
\end{figure}

Figure~\ref{lc_331} shows one of the light curves selected for our blind
test that is affected by a clear jump (top panel) and other less evident discontinuities. 
Into this light curve of a $r'=12.8$ G~dwarf, namely LRc01\_E1\_0331, transits
of a super-Earth with $R_{\rm p}=2.14~R_{\oplus}$ and $P=9.8$~days were inserted. 
The red arrow indicates the discontinuity identified by our pipeline. In the middle panel,
the high- and low-pass filtered light curve (\ref{lam_pipeline}.b-c-d) is displayed. 
An artificial dip followed by a jump is seen at the time of the hot pixel 
as a result of the filtering process. Other artificial flux drops caused by 
small undetected discontinuities are also discernible. Even though the low-pass filter 
reduces the RMS of the light curve, the individual planetary transits are not 
visible in the light curve. 
The artificial dip produced by the hot pixel identified by our pipeline is corrected after
step \ref{lam_pipeline}.e (bottom panel). 
Despite several artificial dips, in this case the transit signal is easily 
detected: the highest peak at the orbital frequency and 
those corresponding to its harmonics
in the BLS spectrum are shown in Fig.~\ref{bls_331}.

f) Transits are searched for by means of the BLS algorithm 
with the directional correction (cf. \citealt{Tingley03}). 
The period search was carried out from $P_{\rm min}=0.4$ 
up to $P_{\rm max}=30$~days. 
The fractionary length of the transit was varied between 
0.006 and 0.09, adopting a number of phase bins $n_{ \rm bins}=240$ 
\citep{Kovacsetal02}.
The frequency sampling $\delta \nu$ was optimised 
according to the criterion given by \citet{Schwarz06}: 
$\delta\nu=1/(P_{\rm max} \cdot n_{\rm bins})$. This criterion ensures that the 
simulated planets do not remain undetected because of
an inappropriate frequency sampling.

g) The light curve phase-folded at the period and epoch found by the
BLS (Fig.~\ref{bls_331}, bottom panel) is visually inspected if events with signal detection efficiency 
greater or equal than 6 have been detected. Events below this threshold are 
generated by pure noise with very high probability \citep{Kovacsetal02}.
In addition, a visual inspection of the original light curve 
filtered from outliers (Sect.~\ref{lam_pipeline}.a) is fundamentally important 
for checking whether a given transit-like feature could be caused by undetected hot pixels.
The visual inspection checks that there are no hot pixels
at any epochs of the detected transits  or,
if any are found, that after their removal, the planetary transit is still detectable and visible
in the folded light curve. A complementary method is checking possible
depth differences when folding the filtered light curve at twice the period found 
by the BLS. Indeed, significant depth differences can arise when the transit-like 
feature is produced by a discontinuity in the light curve.

h) The filtering of an instrumental effect due to the perturbation associated with the 
Earth's rotation at the frequency of one day$^{-1}$ and its harmonics is carried out. 
This effect is caused by variations of the 
Earth's infrared emission that produce temperature changes, even though 
the way in which these variations affect the onboard instrumentation is 
not well understood yet. 
Such a perturbation gives rise to quasi-sinusoidal signals with a period
of 12, 24, or 36~h, affecting all CoRoT light curves, although not in the same
way: there is probably a dependence on the position of the star on the 
CCD. 

\begin{figure}[t!]
\centering
\vspace{-1cm}
\includegraphics[width=10cm, angle=180]{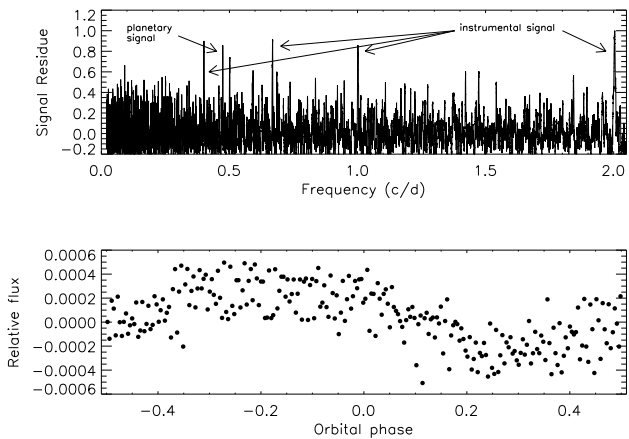}
\vspace{-0.5cm}
\caption{\emph{Top panel}: BLS spectrum of the filtered light curve 
LRa01\_E1\_2996, containing simulated transits by a super-Earth with 
orbital period $P=2.1$~days. The 12~h instrumental effect 
overcomes the planetary signal, which impedes its detection.
\emph{Bottom panel}: phase-folded light curve 
at the period and epoch found by the
BLS, displaying the quasi-sinusoidal variation caused by the
Earth's perturbation (see text for more details).
}
\label{bls_2996_initial}
\end{figure}

\begin{figure}[t!]
\centering
\vspace{-1cm}
\includegraphics[width=10cm, angle=180]{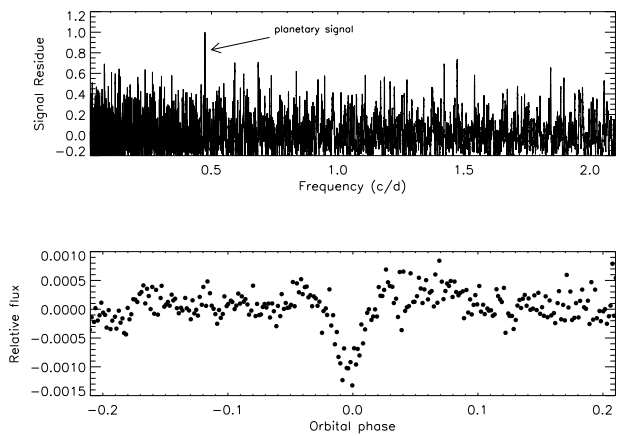}
\vspace{-0.5cm}
\caption{\emph{Top panel}: the same as Fig.~\ref{bls_2996_initial} after 
filtering the 12~h instrumental signal (see Sect.~\ref{lam_pipeline}.h).
The planetary signal is now correctly detected as the most significant 
peak. 
\emph{Bottom panel}: phase-folded light curve 
at the period and epoch found by the
BLS, distinctly showing the transit signature.
}
\label{bls_2996_final}
\end{figure}

Mostly, the Earth's perturbation can be neglected and does not affect 
the transit detection. In a few cases, however, it does have an effect and may lead
to a missed detection. To show this, we display in the top panel of Fig.~\ref{bls_2996_initial} 
the BLS spectrum of the light curve of 
the $r'=15.0$ star LRa01\_E1\_2996, in which we simulated transits 
of a super-Earth with radius $R_{\rm p}=1.56~R_{\oplus}$ and period $P=2.1$~days. We see that 
the peak related to the 12~h instrumental effect exceeds the power of the transit signal, 
which impedes the transit detection. Indeed, from the visual inspection of the 
phase-folded light curve (Sect.~\ref{lam_pipeline}.g), only
a sinusoidal variation with a peak-to-peak amplitude of 
$\sim 6 \cdot 10^{-4}$ appears (see bottom panel of Fig.~\ref{bls_2996_initial}).
To filter this out, we smoothed the quasi-sinusoid in 
phase to determine its amplitude, period, and initial phase, and subtracted it from 
the initial temporal series processed up to step~e). 
After which, carrying out again the BLS allows us to
detect the transit signal because this signal becomes the dominant 
peak in the BLS spectrum (Fig.~\ref{bls_2996_final}).

\begin{table*}[t!]
\centering
\caption{Percentage of detected simulated planets with orbital 
period $0.7 \leq P \leq 20$~days for different values of planetary radius.}            
\renewcommand{\footnoterule}{}     
\begin{tabular}{ccc}       
\hline\hline                 
Planetary radius & Detection rate for G and K  & Detection rate for G and K \\
$[ R_{\oplus} ]$ & dwarfs with $r' \leq 14.0$ $[ \% ]$ & dwarfs with $r' \leq 15.5$ $[ \% ]$  \\
\hline
1.3-2.0  & $10.7 \pm 1.0$ & $3.1 \pm 0.2$ \\
\hline
2.0-3.0  & $41.7 \pm 1.5$ & $16.8 \pm 0.6 $ \\
\hline 
3.0-4.0 & $76.0 \pm 1.6$ & $44.2 \pm 0.8$ \\
\hline
4.0-5.0 & $89.7 \pm 1.2$  & $66.4 \pm 0.8$ \\
\hline
\hline
2.0-4.0 & $58.9 \pm 1.3$ & $30.6 \pm 0.5$ \\
\hline\hline
\end{tabular}
\label{tab_stat_det}      
\end{table*}

\begin{figure*}[t!]
\vspace{-3.0 cm}
\centering
\includegraphics[width=16.0cm]{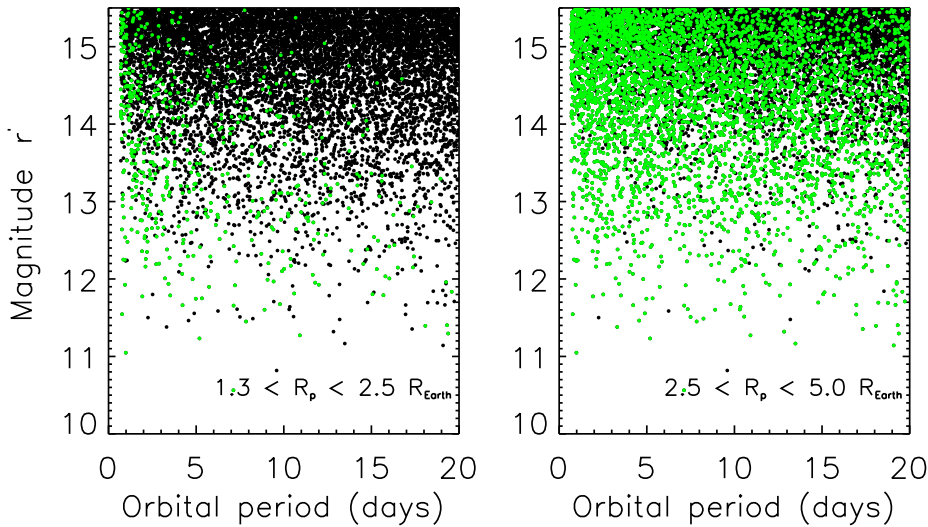}
\vspace{-9.5 cm}
\caption{Distribution of the simulated planets in the $r'$ magnitude vs orbital period
diagram. Green and black dots indicate detected and missed planets, respectively.
\emph{Left}: simulations with $1.3 \leq R_{\rm p} \leq 2.5~R_{\oplus}$. 
\emph{Right}:  simulations with $2.5 < R_{\rm p} \leq 5.0~R_{\oplus}$.
See text for explanation.}
\label{distr_planet}
\end{figure*}

\subsection{Why use a simple median filter?}
Our choice of using a simple median filter to remove stellar variability 
is motivated by two reasons:
the first is the work by \citet{Bonomoetal09}, who showed that iterative
median filters \citep{AigrainIrwin04}, with an appropriate 
choice of the window extension, perform very well for 
detections of planetary transits; the second is
that simple median filters are less sensitive to discontinuities
and/or jumps in the light curves. Indeed, more sophisticated filters, 
such as harmonic fitting \citep{Moutouetal05, BonomoLanza08} 
or wavelet filters \citep{Jenkins02},
give rise to deeper artificial flux drops and oscillations in the vicinity
of an undetected hot pixel. These flux drops, once again, can be 
erroneously identified as transits by the BLS.

\subsection{Optimising the extension of the median filter window}
\label{window_median_filter}
In the same way as in \citet{Bonomoetal09}, to find the optimal window extension of the sliding
median filter, we performed the transit search on the light curves
of a subset of 500 light curves of our blind test by using five different extensions: 
0.3, 0.5, 0.75, 1.0, and 1.25~days.
We found that the maximum detection rate was reached 
with a window extension of 0.5 days, which apparently contradicts
the results of \citet{Bonomoetal09}. These authors showed with extensive simulations
of light curves with solar-like variability and planetary transits that one should use
the longest possible window extension allowed by the stellar activity level
to maximise the transit detection. However, 
their light curves were not affected  by
the instrumental effects that are present in CoRoT data. In other words, the short extension of
0.5~days allows a better filtering of the CoRoT light curves, which are usually 
affected by sudden jumps and jitter noise (Sect.~3.1.e). 
The longer the median filter window, the stronger the filtered light curves
are affected by residual instrumental effects that hamper the transit 
detection. 

One may argue that, on the other hand, for long transits, 
a strong filtering with such a short extension could 
reduce the transit signal. However, 
longer windows did not increase the detection rate in our exercise as we were 
searching for small-size planets around solar-like stars in short-period orbits,
with transit durations of typically less than 5-6~h.

\subsection{The CoRoT alarm mode}
\label{alarm_mode}
The CoRoT alarm mode is a CNES operational task
that is operated at the LAM.
It is responsible for detecting transits in CoRoT light curves 
while observations of a given stellar field are 
ongoing and uses the so-called N1 level data \citep{Suraceetal08}. 
As soon as transits in a light curve are detected
by the alarm mode, the temporal sampling of that light curve 
is changed from 512~s to 32~s. This is very important 
for the transit modelling, which requires the transit ingress and egress 
to be well sampled (e.g., \citealt{Kipping10}), the study of transit timing variations
(e.g., \citealt{Csizmadiaetal10}), and the detection
of the secondary eclipse (e.g., \citealt{Alonsoetal09}). 
Additionally, the alarm mode can trigger 
the ground-based follow-up of the most interesting
planetary candidates as soon as possible, which is required to unveil their 
nature and, if they turn out to be planets, derive the orbital 
parameters. 
If a planetary candidate is not  
detected by the alarm mode, it can be discovered 
by the other detection algorithms of the CoRoT team that search 
for transits on the more precise N2 level data \citep{Cabreraetal09}. 
However, in this case, the ground-based follow-up 
has to wait for about six months
before the target becomes observable again.

With the exception of CoRoT-21b, all CoRoT planets, including the 
super-Earth CoRoT-7b,
were detected by the alarm mode. 
A replacement of the previous variability filter, a gauging filter
developed by \citet{GuisBarge05}, with the one described in 
Sect.~\ref{lam_pipeline}.a-d, 
allowed us to increase the number of planetary 
candidates found by the alarm mode by $\sim 20-30\%$.
This is mainly because the gauging filter 
is very sensitive to discontinuities caused by hot pixels and shows 
that light curve filtering is crucial for detecting
planetary transits, in particular the small-size ones.
CoRoT-21b was missed by the previous version of the
alarm mode pipeline that used the gauging filter. 
The replacement of the variability filter took place 
in February 2009 during the 
long run LRa02. 

The current alarm mode pipeline is a 
simplified version of the one described in Sect.~\ref{lam_pipeline}.
In particular, it performs neither the automatic correction
of light curve discontinuities (Sect.~\ref{lam_pipeline}.e) nor the filtering of 
the instrumental effect mentioned in Sect.~\ref{lam_pipeline}.h. Indeed,
the task of the alarm mode is, in principle, 
the detection of naked-eye transits, hence the 
complete LAM detection pipeline (Sect.~\ref{lam_pipeline}) is used 
only to search for small-size planets with
N2 data. Nevertheless, even with this simplified 
detection pipeline, the alarm mode enables detecting many
small-size planetary candidates with transits embedded in the noise.
As an example, it led to the detection 
of the Neptune CoRoT-24c \citep{Alonsoetalsubm}, with a transit $S/N$ lower
than CoRoT-7b by about a factor of three since it orbits a faint star 
with $r'=15.11$ in almost 12~days.

\section{Results and discussion}
\label{results}
\subsection{CoRoT detection rate of super-Earths and Neptunes}
\label{corot_det_rate}

\noindent
With our LAM transit detection pipeline (Sect.~\ref{lam_pipeline}), we searched 
for the simulated transits blindly, that is, without knowing in advance into which
light curves the transits had been inserted (Sect.~\ref{bt5}). Figure~\ref{distr_planet}
shows our results in terms of detected planets in the 
$r'$ magnitude vs orbital period diagram. The green and black dots indicate the 
detected and undetected events, respectively. 
We considered an event as ``detected'' when the BLS algorithm correctly found 
the period of the simulated transits and the signal detection efficiency was $\geq 6$.
In the left panel of Fig.~\ref{distr_planet}, we clearly
notice that only few planets with radius $R_{\rm p} \leq 2.5~R_{\oplus}$, i.e. 6\%,
were correctly found. The vast majority of them orbit relatively bright stars and/or 
have short orbital periods, i.e. $P<5$~days. In contrast, the right panel shows
that about 50\% of the simulated planets with larger radius, 
$2.5 <R_{\rm p} \leq 5.0~R_{\oplus}$, were correctly detected. In the latter case, the majority of
the missed events are the simulated transits with relatively long orbital periods, $P>13$~days,
which were inserted in faint stars with $r' >14.5$ (see Fig.~\ref{distr_planet}, right panel).

Table~\ref{tab_stat_det} reports the percentage of detections of the simulated transits
with orbital period $0.7 \leq P \leq 20$~days for different intervals of planetary radii. 
The associated errors were estimated with binomial statistics.
Below 2~$R_{\oplus}$, the detection rate $\xi_{\rm CoRoT}$ is very low, 
$\xi_{\rm CoRoT} \sim 3 \%$ 
by taking all the stars of our blind tests into account, 
and $\xi_{\rm CoRoT} \sim 11 \%$ by considering only those brighter
than $r' = 14.0$. For the biggest planets with 
$4.0 \leq R_{\rm p} \leq 5.0~R_{\oplus}$, $\xi_{\rm CoRoT} \sim 2/3$ (9/10) 
for $r' \leq 15.5$ ($\leq 14.0$). By taking small-size planets
with radii between 2 and 4~$R_{\oplus}$ into account, which is the smallest range considered by
\citet{Howardetal12} to determine the planet occurrence rate,
our detection rate is $30.6 \pm 0.5\%$ and $58.9 \pm 1.3\%$ for stars brighter than $r' = 15.5$ and 14.0, 
respectively. For this range of planetary radius, Table~\ref{tab_stat_period} lists 
the fraction of detected planets in eight bins of orbital period, along with their own 
uncertainties estimated, once again, with binomial statistics.

\begin{table}[h!]
\centering
\caption{Percentage of detected simulated planets with radius 
$2 \leq R_{\rm p} \leq 4~R_{\oplus}$ as a function of orbital period.}            
\renewcommand{\footnoterule}{}     
\begin{tabular}{cc}       
\hline\hline                 
Orbital period & Detection rate for G and K   \\
$[\rm days ]$ & dwarfs with $r' \leq 15.5$ $[ \% ]$   \\
\hline
0.7-2.5  & $65 \pm 2$  \\
\hline
2.5-5.0  & $46 \pm 2$ \\
\hline 
5.0-7.5 & $36 \pm 2$ \\
\hline
7.5-10.0 & $30 \pm 1$  \\
\hline
10.0-12.5  & $24 \pm 1$ \\
\hline
12.5-15.0  & $22 \pm 1$ \\
\hline
15.0-17.5  & $18 \pm 1$ \\
\hline
17.5-20.0  & $15 \pm 1$ \\
\hline

\hline\hline
\end{tabular}
\label{tab_stat_period}      
\end{table}

\begin{figure}[h!]
\centering
\vspace{-0.5cm}
\includegraphics[width=7.0cm, angle=-90]{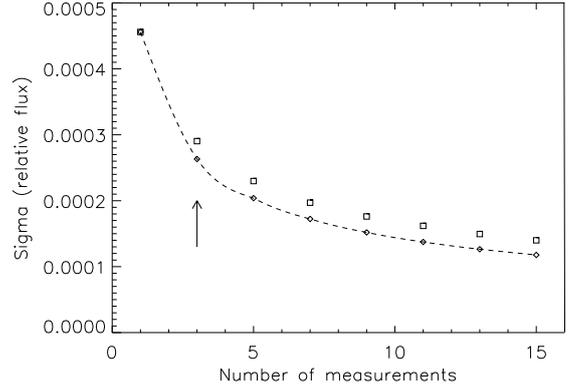}
\vspace{-0.5cm}
\caption{Measured noise (square) and expected white noise (diamond) of a typical 
CoRoT light curve used in our blind test as 
a function of the number of photometric measurements sampled at 512~s.
In this case, the highest difference is found at 3 measurements, as indicated by 
the arrow. 
}
\label{corr_noise}
\end{figure}

For each simulated transiting planet we computed
a posteriori the signal-to-noise ratio 
as $S/N= \it (\delta / \sigma) \cdot \sqrt{n_{\rm tr}}$, where
$\delta$ is the transit depth, $n_{\rm tr}$ the number of data points inside the transit
bottom part, and $\sigma$ the total noise of the 512~s sampled light curves. 
The latter is estimated as 
$\sigma=\sqrt{\sigma_{\rm w}^{2}+\sigma_{\rm r}^{2}}$, where $\sigma_{\rm w}$ is the 
point-to-point scatter of the out-of-transit light curve 
and $\sigma_{\rm r}$ its correlated noise (e.g., \citealt{Pontetal06}). 
$\sigma_{\rm w}$ was evaluated in a robust way as $\sigma_{\rm w}=1.4826 \cdot \rm MAD$,
where MAD stands for median absolute deviation, to exclude 
possible outliers, after removing low-frequency variations on 
time scales longer than 12~hours. 
$\sigma_{\rm r}$ was estimated by 
comparing the expected white noise at seven different time scales,
from 25~min (3~data points) up to 2.1~h (15~data points), with the standard deviation of the 
sliding average over the same time scales (evaluated, once again, in a robust way). 
Since the white noise decreases with the square root 
of the number of measurements,
the maximum of the quadrature differences between 
the expected white noise and the measured noise, computed for 
the seven temporal windows, was set to $\sigma_{\rm r}$. 
Fig.~\ref{corr_noise} is an example of the estimated noise in 
a typical CoRoT light curve used in our blind test.
In agreement with estimates of the correlated noise in CoRoT light curves
by \citet{Aigrainetal09}, we found that its contribution is in 
general low, $\sim 1/4$ of the total noise. 

\begin{figure*}[t!]
\vspace{-3.0 cm}
\centering
\includegraphics[width=16.0cm]{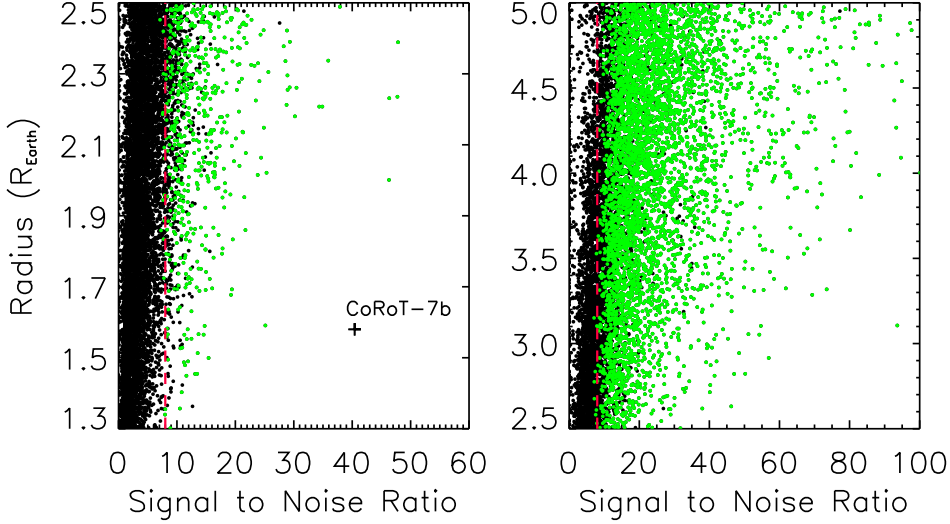}
\vspace{-9.5 cm}
\caption{Radius of the simulated planets as a function of the
$S/N$ of their transits. As in Fig.~\ref{distr_planet}, green and black dots 
indicate the identified and missed planets, respectively. 
The vast majority of the discovered transits have $S/N > 8$ (red dashed line). 
Note that the x-scale in the left and right panels is different.
See text for discussion about the transits with $S/N \ge 15$ that should be easily 
detectable but were missed in our blind test.}
\label{radius_SNR}
\end{figure*}

Figure~\ref{radius_SNR} displays the radius of the simulated 
planets in terrestrial units as a function of the signal-to-noise 
ratio of their transits. As in Fig.~\ref{distr_planet},
green dots indicate the correctly identified planets while 
the black dots the missed ones. 
Figure~\ref{radius_SNR} shows that the vast 
majority of detected transits have $S/N > 8$ and that 
CoRoT is able to detect transiting 
small-size planets with an $S/N$ much lower than CoRoT-7b, i.e. 
$\sim~1/4 \cdot S/N_{\rm CoRoT-7b}$.

\subsection{Obstacles in detecting small-size planets in CoRoT data}
\label{det_obstacles}
The vast majority of undetected planets in our blind test were missed because 
their transits have a low $S/N$. However, some events with a relatively high $S/N$, 
which should have been quite easily identified, were also missed. This motivated us 
to investigate additional obstacles that may contribute to a missed detection.
For this reason, we carefully looked at the few events 
with $S/N > 15$ that went undetected 
(see Fig.~\ref{radius_SNR}). They represent only 5\% of the
total number of simulated planets with $S/N > 15$
because 95\% of them were correctly found.

The first obstacle has been mentioned already  
(Sect.~\ref{lam_pipeline}.e, h) and consists of light 
curve discontinuities mainly caused by hot pixels or 
pointing displacements that were not identified 
by our pipeline because of their low amplitude. 
Another obstacle is the stellar variability in the case of 
fast rotators with rotation period $P_{\rm rot} < 2-3$~days.
In this case, the variability is much stronger than the transit signal. 
Even a very strong filtering is unable
to reveal the transits because it simultaneously reduces their depth,
which prevents us from detecting them.
We found that $\sim 70 \%$ of the missed transits with $S/N > 15$ went
undetected by our pipeline owing to unidentified, hence 
uncorrected light curve discontinuities, and $\sim 30 \%$ because of stellar variability.

While such obstacles are not a major concern for events with 
$S/N > 15$ since, as already mentioned, only 5\% of such
planets were not detected, 
they certainly have a stronger impact on signals near the detection
threshold, i.e. $S/N \sim 10$. Therefore, we suspect they are very likely
responsible for the 20\% of small planets with $S/N \geq 10$ 
that were not found by our pipeline.

Finally, we recall that the instrumental signal caused by Earth's perturbation
on 12 or 24~h time scale  can in some cases also impede the detection
of small-size planets, if it is not adequately filtered. 
However, this filtering is implemented in our automatic pipeline 
(Sect.~\ref{lam_pipeline}.h).

\subsection{Comparison between discovered and 
expected CoRoT planets with $R_{\rm p}=2-4~R_{\oplus}$.}
\label{comp_kepler}
After deriving the CoRoT detection rate of small-size planets
from our blind test, we are now able to estimate the expected number of
CoRoT planets with radius between 2 and 4~$R_{\oplus}$ 
from the statistics provided by the distribution of the Kepler planetary candidates.
Assuming that these candidates are real planets \citep{MortonJohnson11}, 
\citet{Howardetal12} found, by studying their occurrence rate around 
G and K dwarfs after correcting for the alignment probability, 
that ``the planet occurrence increases substantially with decreasing
planet radius and increasing orbital period'' (see their Figures~5 and 6). 
They empirically determined the planet occurrence as a function of 
planetary radius and orbital period up to
50 days (cf. their Equations~4 and 8). 
In the remainder of the paper we consider only planets 
with $2 \leq R_{\rm p} \leq 4~R_{\oplus}$  and 
$1.2 \leq P \leq 17.0$~days to 
choose the same bins of planetary radius and orbital period as \citet{Howardetal12} 
(see their Figures~4 and 7). Integrating between 1.2 and 17.0~days gives 
an occurrence rate of $5.5 \pm 0.5 \%$. 

The expected number of CoRoT planets with radii of $2-4~R_{\oplus}$
and orbital periods up to 17~days can be estimated as

\small

\begin{eqnarray}
\label{pl_corot}
n_{\rm CoRoT[2-4~R_{\oplus}]}= n_{\rm \star [GV-KV]} \int^{4}_{2} \int^{17.0}_{1.2}  
f(R, P) & p_{\rm tr}(P) & \xi_{\rm CoRoT}(R, P) \nonumber \\
& & dR~dP~,
\end{eqnarray}

\normalsize

\noindent
where $n_{\rm \star [GV-KV]}$ is the number of G and K 
main-sequence stars with $r' \leq 15.5$ in the six 
CoRoT long runs used in our blind test  (Sect.~\ref{bt5}),  
i.e. those lasting more than 110~days, hence
more suited to the search for small-size planets;
$f(R, P)$ is the frequency of planets as a function of planetary radius
and orbital period determined by \citet{Howardetal12}; 
$p_{\rm tr}(P)=[(4 \pi^2 R_{\star}^{3})/(GM_{\star}P^2)]^{1/3}$ is the transit probability
assuming for simplicity $R_{\star}=R_{\odot}$, $M_{\star}=M_{\odot}$, and 
circular orbits; $\xi_{\rm CoRoT}(R, P)$ is the CoRoT detection
rate, which was derived from our blind test
and is a function of planetary radius and orbital period. 
The integral $\int^{4}_{2} \int^{17.0}_{1.2}  f(R, P)~p_{\rm tr}(P)~dR~dP$ 
represents the fraction of transiting Neptune-size planets
with $P \leq 17$~days according to the Kepler planet occurrence (we recall that
20~days is the maximum allowed period in our blind test). The integral in
Eq.~\ref{pl_corot}, by containing also the factor $\xi_{\rm CoRoT}(R, P)$, 
gives the percentage of such planets detectable by CoRoT. 
Multiplying the integral by $n_{\rm \star [GV-KV]}$ 
simply allows one to estimate the number of Neptune-size planets that CoRoT 
should have discovered ($n_{\rm CoRoT[2-4~R_{\oplus}]}$, see Eq.~\ref{pl_corot}). 
This was evaluated by computing the integrand function
and multiplying it by $n_{\rm \star [GV-KV]}$
for each of the ``cells'' that range from 2 to 4 $R_{\oplus}$ and from 1.2 
to 17.0~days in Fig.~4 of \citet{Howardetal12} (see Table~\ref{tab_stat_integral}).  
The values calculated for each cell were then added.

Because G and K dwarfs are underestimated in the Exodat photometric classification, 
as pointed out by \citet{Gazzanoetal10}, 
$n_{\rm \star [GV-KV]}$ was increased by a factor 1.7:  
$n_{\rm \star [GV-KV]}=\rm 9526~x~1.7=16194$ (see Sect.~\ref{bt5} and 
Fig.~8 in \citealt{Gazzanoetal10}). Moreover, a conservative arbitrary error of 30\% 
on this quantity was considered because the work by \citet{Gazzanoetal10} 
inevitably relies on a limited number of stars, $\sim 1\,000$. 
In this way, the number of CoRoT planets $n_{\rm CoRoT [2-4~R_{\oplus}]}$ 
expected from the planet occurrence rate determined by \citet{Howardetal12} 
was estimated to be $12 \pm 2$ (Eq.~\ref{pl_corot}). 
The error was evaluated with a Monte Carlo method 
and is related to the uncertainties on $n_{\rm \star [GV-KV]}$,  
$f(R, P)$, and $\xi_{\rm CoRoT}(R, P)$. The error on $f(R, P)$ 
was computed with binomial statistics 
in the same way as described in \citet{Howardetal12}, Sect.~3 
(see Table~\ref{tab_stat_integral}).

\begin{table*}
\centering
\caption{Kepler planet frequency $f(R, P)$, CoRoT detection $\xi_{\rm CoRoT}(R, P)$ rate, and number of  
expected CoRoT Neptunes $n_{\rm CoRoT [2-4~R_{\oplus}]}$ for each cell of $R_{p}$ and $P$ as in Fig.~4 of \citet{Howardetal12}.}            
\renewcommand{\footnoterule}{}     
\begin{tabular}{|ll|ccc|}       
\hline              
Planetary radius  & Orbital period & $f(R, P)$ & $\xi_{\rm CoRoT}(R, P)$ & $n_{\rm CoRoT [2-4~R_{\oplus}]}$ \\
$[R_{\oplus}]$& $[\rm days ]$ & $[ \% ]$ & $[ \% ]$ &    \\
\hline
2.0-2.8 & 1.2-2.0 & $0.037 \pm 0.021$ & $44 \pm 6$ & $0.45 \pm 0.31$  \\
\hline
2.0-2.8 & 2.0-3.4 & $0.11 \pm 0.04$ & $28 \pm 4$ & $0.61 \pm 0.31$  \\
\hline 
2.8-4.0 & 2.0-3.4 & $0.035 \pm 0.020$ & $71 \pm 3$ & $0.49 \pm 0.33$  \\
\hline
2.0-2.8 & 3.4-5.9 & $0.51 \pm 0.11$ & $19 \pm 1$ & $1.31 \pm 0.50$  \\
\hline
2.8-4.0 & 3.4-5.9 & $0.18 \pm 0.06$ & $60 \pm 2$ & $1.48 \pm 0.68$  \\
\hline
2.0-2.8 & 5.9-10.0 & $1.0 \pm 0.2$ & $14 \pm 1$ & $1.34 \pm 0.50$  \\
\hline
2.8-4.0 & 5.9-10.0 & $0.62 \pm 0.13$ & $44 \pm 2$ & $2.62 \pm 0.98$  \\
\hline
2.0-2.8 & 10.0-17.0 & $1.9 \pm 0.3$ & $7 \pm 1$ & $0.85 \pm 0.30$  \\
\hline
2.8-4.0 & 10.0-17.0 & $1.1 \pm 0.2$ & $31 \pm 1$ & $2.43 \pm 0.83$  \\
\hline
\end{tabular}
\label{tab_stat_integral}      
\end{table*}

We point out that our estimate of $n_{\rm CoRoT[2-4~R_{\oplus}]}$ might
be slightly underestimated because in our simulations the stellar radius 
and mass were fixed to those of a G2V or K2V star for the G and 
K dwarfs, respectively\footnote{No information on stellar radii is available in the Exodat catalogue and, in the 
absence of such an information, we preferred to be conservative.}.
G stars of later spectral types would give rise to deeper, hence more easily detectable
transits. Therefore, it is likely that our detection rate $\xi_{\rm CoRoT}(R, P)$ is moderately underestimated.
However, this effect is partly compensated for by the slightly lower transit probability 
for such stars.

After estimating $n_{\rm CoRoT[2-4~R_{\oplus}]}$, a comparison 
with the CoRoT Neptunes and Neptune-size planetary candidates is interesting. For that 
purpose, the N2 light curves of the six observational runs considered for our 
blind test were analysed with the LAM pipeline (Sect.~\ref{lam_pipeline}). All small-size candidates 
previously detected by the alarm mode (Sect.~\ref{alarm_mode}) and the other six detection algorithms
used by the CoRoT team, including those discovered by \citet{Ofiretal10}, were 
correctly found. The CoRoT Neptune-size planetary candidates,
regardless of their priority and stellar classification,
constitute a substantial fraction of the total number of candidates 
(Deleuil et al., in preparation; see, e.g., Fig.~14 in \citealt{Caroneetal12} ).

Since the statistics provided by \citet{Howardetal12} does not 
take planets smaller than $2~R_{\oplus}$ into account 
because of the current incompleteness of Kepler candidates,
we excluded CoRoT-7b from this comparison. By considering
only the range between 2 and 4~$R_{\oplus}$, we discarded also
CoRoT-8b \citep{Bordeetal10} and CoRoT-24c \citep{Alonsoetalsubm} 
because all of them have radii larger than 4~$R_{\oplus}$.  
The only CoRoT validated planet in this range is CoRoT-24b, 
which has a radius of $3.7 \pm 0.4~R_{\oplus}$ and an 
orbital period of 5.11~days \citep{Alonsoetalsubm}. 
We also checked the CoRoT Neptune-size candidates that
were not identified as false positives from the light curve
analysis and follow-up ground-based observations \citep{Deegetal09, Moutouetal09}.
The nature of these candidates is unknown, however, because
no radial-velocity variation was detected with the HARPS
spectrograph given that they orbit faint stars ($r' \gtrsim 13$).
Therefore, they could be either real planets,
in which case only an upper limit can be put on their mass,
or blended eclipsing binaries not identifiable from
the CoRoT light curves.
For uniformity with our study, we selected the Neptune-size
candidates orbiting stars with $r' \leq 15.5$ that 
are classified as G or K dwarfs by Exodat,
in the long runs LRa01, LRa02, LRa03, LRa04, LRc01, 
and LRc02. 
We ended up with three candidates and also added two more candidates
whose stars were misclassified as giants by the
Exodat photometric classification but were recognised to be G8V 
and K3V stars, respectively, thanks to low-resolution 
reconnaissance spectroscopy. 
However, we point out that the transits of two of them are V-shaped, which
would indicate they are more likely blends.

In the most favourable case that all these candidates are real planets, which cannot be 
stated with certainty as they have not been properly validated, 
there would be six CoRoT Neptunes. This therefore disagrees with the 
expectation from the Kepler mission, i.e. 
$n_{\rm CoRoT[2-4~R_{\oplus}]}=12 \pm 2$, at  3~$\sigma$. 
The disagreement would obviously get worse if a significant fraction of the Neptune candidates 
were actually blends, up to 5~$\sigma$
for a blend fraction of 100\%.

Such a disagreement may be due to several reasons. First of all,
the uncertainty on $f(R, P)$ might be underestimated 
because the errors on the size of the Kepler
planetary candidates were not taken into consideration by
\citet{Howardetal12}. Such errors are related to the uncertainties
on the stellar radii in the KIC catalogue which are
typically $\sim 30\%$ \citep{Brownetal11}.
Secondly, the false-positive probability (FPP) of Kepler small-size planetary candidates in
single systems might be higher than predicted by \citet{MortonJohnson11}.
This has been suggested by \citet{Colonetal12}, who followed-up  
four Kepler candidates with radii smaller than $5~R_{\oplus}$ and orbital periods 
less than 6 days with multicolour photometry. Half of them turned out to be false 
positives \citep{Colonetal12}. However, Spitzer observations of a 
bigger sample of Kepler candidates would indicate, on the contrary, a low FPP
\citep{Desertetal12}.
An FPP higher than estimated by \citet{MortonJohnson11} 
may have led \citet{Howardetal12} to overestimate the planet occurrence
of small-size planets, although at least the Kepler planetary candidates in
multiple systems are in all likelihood real planets \citep{Lathametal11, Lissaueretal12}.
An alternative reason might be that a wrong value was adopted 
for $n_{\rm \star [GV-KV]}$ but the latter should be sufficiently accurate
after correcting for the systematic effects of the photometric stellar 
classification and adding an arbitrary error of $30\%$.
Lastly, the disagreement could also be related to the different 
stellar populations observed by CoRoT and Kepler.
In any case, this disagreement needs to be investigated in more detail 
in subsequent papers as its explanation goes beyond the scope of the present work.

Despite the above-mentioned disagreement and the 
fact that the exact number of CoRoT Neptunes cannot be
precisely determined for the moment, we point out that their noticeable deficiency
seems to agree with the general trend found in Kepler data,
i.e. that the frequency of small-size planets increases with increasing orbital 
periods and decreasing planet radii \citep{Howardetal12}.
If transiting planets with $2.0 \leq R_{\rm p} \leq 4.0~R_{\oplus}$ were numerous 
at short-orbital period, a significant fraction of them would 
have been detected by CoRoT, up to $59\%$ ($31\%$) for G and K 
dwarfs brighter than $r' = 14.0$ (15.5)
(Table~\ref{tab_stat_det}).

\section{Summary and prospects}
We performed a blind test that consisted in simulating transits of small-size planets, 
super-Earths and Neptunes, in real CoRoT light curves of G and K dwarfs, 
and searching for them blindly. 
In this way, we investigated the capability of CoRoT 
of detecting small-size planets in short-period orbits. In particular, we showed that CoRoT
can detect small planets with an $S/N$ as low as $\sim 1/4$ that of CoRoT-7b.
We found that the CoRoT detection rate is 59\% (31\%) 
for planets with radii between 2 and $4~R_{\oplus}$ orbiting stars with
$r' \leq 14.0$ ($\leq 15.5$) in $P \leq 20$~days. 
The vast majority of missed planets went undetected because of 
a low $S/N$ of their transits. However, in some cases, additional instrumental or 
astrophysical noise may also prevent transits with 
relatively high $S/N$, i.e. $S/N \geq 10$, from being revealed.
This noise is mainly caused by uncorrected light curve discontinuities 
produced by hot pixels and pointing displacements, or 
short-term ($<2-3$~days) stellar variability.

By properly taking the CoRoT detection rate for Neptune-size planets
($2 \leq R_{\rm p} \leq 4~R_{\oplus}$) and the transit probability into account, 
CoRoT should have discovered $12 \pm 2$ Neptunes 
with $P \leq 17$~days around G and K dwarfs of six observational runs, according to the Kepler planet 
occurrence rate \citep{Howardetal12}. However, between one and six
CoRoT Neptunes were found in these runs, 
depending on the fraction of Neptune-size candidates
that are actually blends.
The reason of this disagreement still needs to be investigated. In any case, 
by noticing the scarcity of close-in Neptune-size planets found by CoRoT, 
our results seem to indirectly support the reality of the trend found in Kepler data, 
i.e. that the frequency of small-size planets increases with increasing orbital 
periods and decreasing planet radii \citep{Howardetal12}. 

At the end of the CoRoT mission, repeating this comparison 
by including the next CoRoT long runs and new detected or 
validated CoRoT Neptunes will certainly prove instructive. Moreover, 
the radial-velocity follow-up of Kepler 
small-size planetary candidates with the high-resolution spectrograph HARPS-N, 
mounted at the Telescopio Nazionale Galileo (La Palma, Spain), 
will also unveil their FPP, while determining the mass of several of them. 
The FPP of Kepler candidates is of fundamental importance for studies on the planet occurrence 
rate and could be then compared with the theoretical estimate of \citet{MortonJohnson11}
adopted by \citet{Howardetal12}.

\begin{acknowledgements}
A. S. Bonomo gratefully acknowledges support through INAF/HARPS-N fellowship
and CNES grant.  A. S. Bonomo, C. Moutou, M. Deleuil, F. Bouchy, and A. Santerne 
acknowledge support from the ``Programme National de Plan\'etologie'' (PNP) 
of CNRS/INSU and the French National Research Agency (ANR-08-JCJC-0102-01).
The authors wish to thank R. F. D\'iaz for interesting and useful discussions.
\end{acknowledgements}

~\\
~\\

\end{document}